\documentclass[11pt]{article}
\usepackage{graphicx}
\usepackage{amsmath}

\newcommand\pubnumber{DPF2015-254}
\newcommand\pubdate{\today}

\def\omegapinc{\textsuperscript{1}Omega-P, Inc, New Haven, CT 06511, USA; \textsuperscript{2}Yale University, New Haven, CT, USA}
\def\support{\footnote{This research was supported by the DoE Phase I grant DE-SC0011347}}

\def\Title#1{\begin{center} {\Large #1 } \end{center}}
\def\Author#1{\begin{center}{ \sc #1} \end{center}}
\def\Address#1{\begin{center}{ \it #1} \end{center}}

\newcommand\pubblock{\rightline{\begin{tabular}{l} \pubnumber\\
         \pubdate  \end{tabular}}}
\newenvironment{Abstract}{\begin{quotation}  }{\end{quotation}}
\newenvironment{Presented}{\begin{quotation} \begin{center} 
             PRESENTED AT\end{center}\bigskip 
      \begin{center}\begin{large}}{\end{large}\end{center} \end{quotation}}

\def\beq{\begin{equation}}
\def\eeq#1{\label{#1}\end{equation}}
\def\eeqn{\end{equation}}

\def\beqa{\begin{eqnarray}}
\def\eeqa#1{\label{#1}\end{eqnarray}}
\def\eeqan{\end{eqnarray}}

\let\bar=\overbar

\def\Dslash{\not{\hbox{\kern-4pt $D$}}}
\def\dslash{\not{\hbox{\kern-2pt $\del$}}}

\def\msb{{\bar{\ssstyle M \kern -1pt S}}}

\begin{document}
\begin{titlepage}
\pubblock

\vfill
\Title{DIAGNOSTIC AND DETECTORS FOR CHARGING AND DAMAGE OF DIELECTRICS IN HIGH-GRADIENT ACCELERATORS\support
}
\vfill
\Author{ S.V. Shchelkunov\textsuperscript{1,2} (sergey.shchelkunov@gmail.com),\\ T.C. Marshall\textsuperscript{1}, J.L. Hirshfield\textsuperscript{1}}
\Address{\omegapinc}
\vfill
\begin{Abstract}
This research addresses issues of analysis and mitigation of high repetition rate effects in Dielectric Wakefield Accelerators, and more specifically, to study charging rate and charge distribution inside a thin walled dielectric wakefield accelerator as may occur from a passing charge bunch, and the physics of conductivity and discharge phenomena at the dielectric wall.  The issue is the role played by the beam halo and intense wakefields in charging of the dielectric surface, possibly leading to undesired deflection of charge bunches and degradation of the dielectric material. During initial stage of development, microwave apparatus was built and signal processing was developed for observing time-dependent charging of dielectric surfaces and/or plasmas located on or near the inner surface of a thin-wall hollow dielectric tube that fits within a microwave cavity resonator. Three frequencies were employed to improve the data handling rate and the signal-to-noise. The test and performance results for a plasma test case will be presented; in particular, the performance of the test unit showed capability to detect small changes $\sim$~0.1\% of a dielectric constant, which would correspond to the scraping-off of only 0.3 nC to the walls of the dielectric liner from the passing charge bunch.
\end{Abstract}
\vfill
\begin{Presented}
DPF 2015\\
The Meeting of the American Physical Society\\
Division of Particles and Fields\\
Ann Arbor, Michigan, August 4--8, 2015\\
\end{Presented}
\vfill
\end{titlepage}
\def\thefootnote{\fnsymbol{footnote}}
\setcounter{footnote}{0}

\section{Introduction}

Dielectric Wakefield Accelerators (DWA) have many favorable properties, among them, high gradient acceleration of electron or positron bunches in a simple smooth-bore structure.  While the simplest model, a structure in which a single drive bunch sets up a wakefield in a dielectric-lined hollow cylindrical tube and is followed by a witness bunch which is accelerated by the wakefield, has problems with breakup of a high charge drive bunch~\cite{ref1} and low transformer ratio~\cite{ref2}, more complicated structures, e.g. a two-channel coaxial configurations~\cite{ref3} offer solutions to these difficulties.  In the past, there has been experience with suitable dielectrics~\cite{ref4} but there has been no systematic investigation about the suitability of certain dielectrics for application in the DWA.  This we deem to be a serious omission, and it is the aim of the present investigation to develop apparatus and methods to obtain evidence for dielectric damage that can occur in an accelerator environment.
Herein, we describe the performance of diagnostic hardware and procedures that were developed for the study of charging effects, damage, and conductivity changes resulting from the passage of one or more charged bunches, e.g. because of the deposition of the halo particles on the dielectric material, or effects caused by intense wakefields – as set up by the drive bunches - on the dielectric material.

\section{Approach and method}
The following summarizes our approach and method, namely:
 
(step 1) Surround the dielectric material by a microwave cavity. The cavity is manufactured to be fitted around the dielectric thin-wall tubing. The working mode TM$ _{010} $ of the cavity is chosen to be in a low GHz frequency range (e.g. \textless 20 GHz). The fundamental mode can be recommended since it is situated far away from the nearest higher-order mode.   

(step 2) Measure the changes in transmission through the cavity at pre-defined frequencies, f$ _{m} $.The measurements are done in the real time, at (as it will be shown later) micro-second time-scales or above.  The set of frequencies is chosen so that each is relatively close to the resonant frequency, F$_{\xi}$, of the chosen cavity mode. For instance, if the cavity bandwidth is BW$_{\xi}$, the recommended frequency choice is f$_{m}$ = F$_{\xi}$  $\pm$ BW$_{\xi}\cdot N _{m}$/2, where $N _{m}$ = 0, 1, 2,..., etc .

(step 3) Using the above information, infer the changes in the cavity resonant frequency, F, and Q-factor. This is done by application of a set of pre-derived formulas, which will be given below in subsequent sections. At this point one also uses an already developed and tested noise-rejection and filtering algorithm to boost the signal-to-noise ratio.

(step 4)  Find the corresponding changes in the dielectric constant and loss-tan of the dielectric  thin-walled tubing or plasma discharge. To do so, one inputs e.g. the known dielectric material geometrical configuration into HFSS (or similar software), and runs the code to find what the dielectric constant and loss-tan should be at each given time-moment to cause the observed changes in the cavity resonant frequency and Q-factor.  Note that this will allow one to monitor/detect both transient effects as well as permanent ones.

(step 5) Use models to find the corresponding density of free electrons/ charges (n$ _{e} $), collision frequency (F$ _{coll} $), and  the other relevant parameters (if any).

(step 6) Use the data from above to predict changes in the dielectric constant and loss-tan of the dielectric material in GHz- or other-frequency ranges. Again, by the nature of the techniques, both transient changes on the microsecond time scale or greater as well as permanent ones can be inferred.

\section{Description of a proof - of – principle experiment to validate the technique and performance}

For the proof-of-principle experiment – having no access to a suitable beam-line that would provide a dielectric wakefield accelerator (namely a relativistic charge bunch that passes through a hollow dielectric tube) - we decided to apply our technique to the case of a plasma discharge contained within a quartz tube (the dielectric here), since it is well known how it behaves~\cite{ref5},~\cite{ref7}. Figure 1 (left) shows the diagram of the setup. The discharge tube was located along the axis of the cavity that has a resonant frequency F$_{\xi}$ = 2479.157MHz, and Q$_{\xi}$ = 1175 (corresponding to BW$_{\xi}$ = 2.11 MHz); the values are denoted with the subscript $\xi$ to indicate that they are for when there is no discharge or deposited/free electrons present. The changes in transmission were monitored at 3 different frequencies: f$_{1}$ $\approx$ F$_{\xi}$; f$_{2}$ $\approx$ F$_{\xi}$ + (BW$_{\xi}$/2) and f$_{3}$ $\approx$ F$_{\xi}$ + BW$_{\xi}$, (the other choice was to monitor at f$_{1}$ $\approx$ F$_{\xi}$ - (BW$_{\xi}$/2) ; f$_{2}$ $\approx$ F$_{\xi}$  and 	f$_{3}$ $\approx$ F$_{\xi}$ + BW$_{\xi}$/2).
 
\begin{figure}[htb]
\centering
\begin{minipage}{3.75in}
\centering
\includegraphics[width=3.75in]{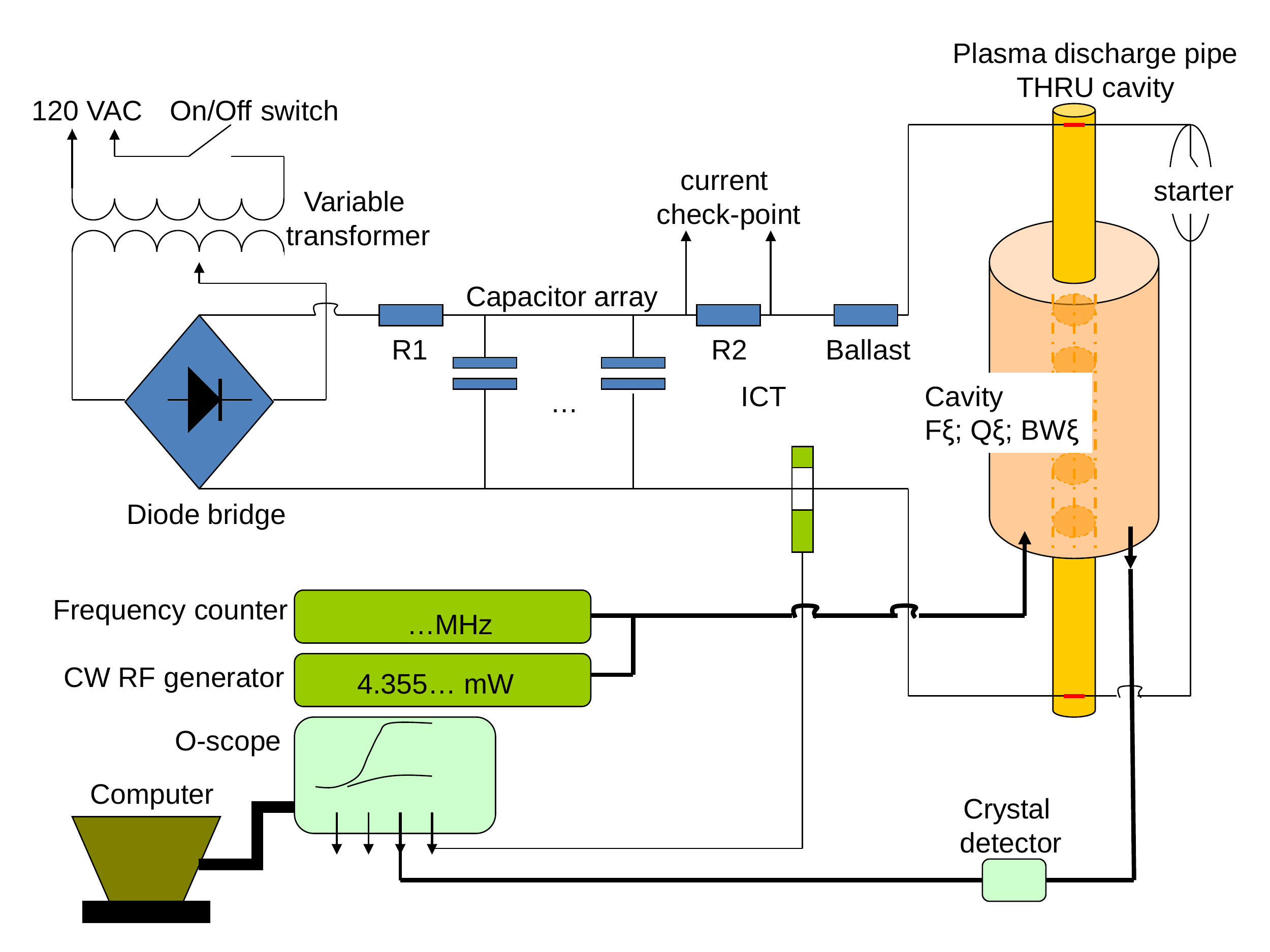}
\end{minipage}
\begin{minipage}{2in}
\centering
\includegraphics[width=2in]{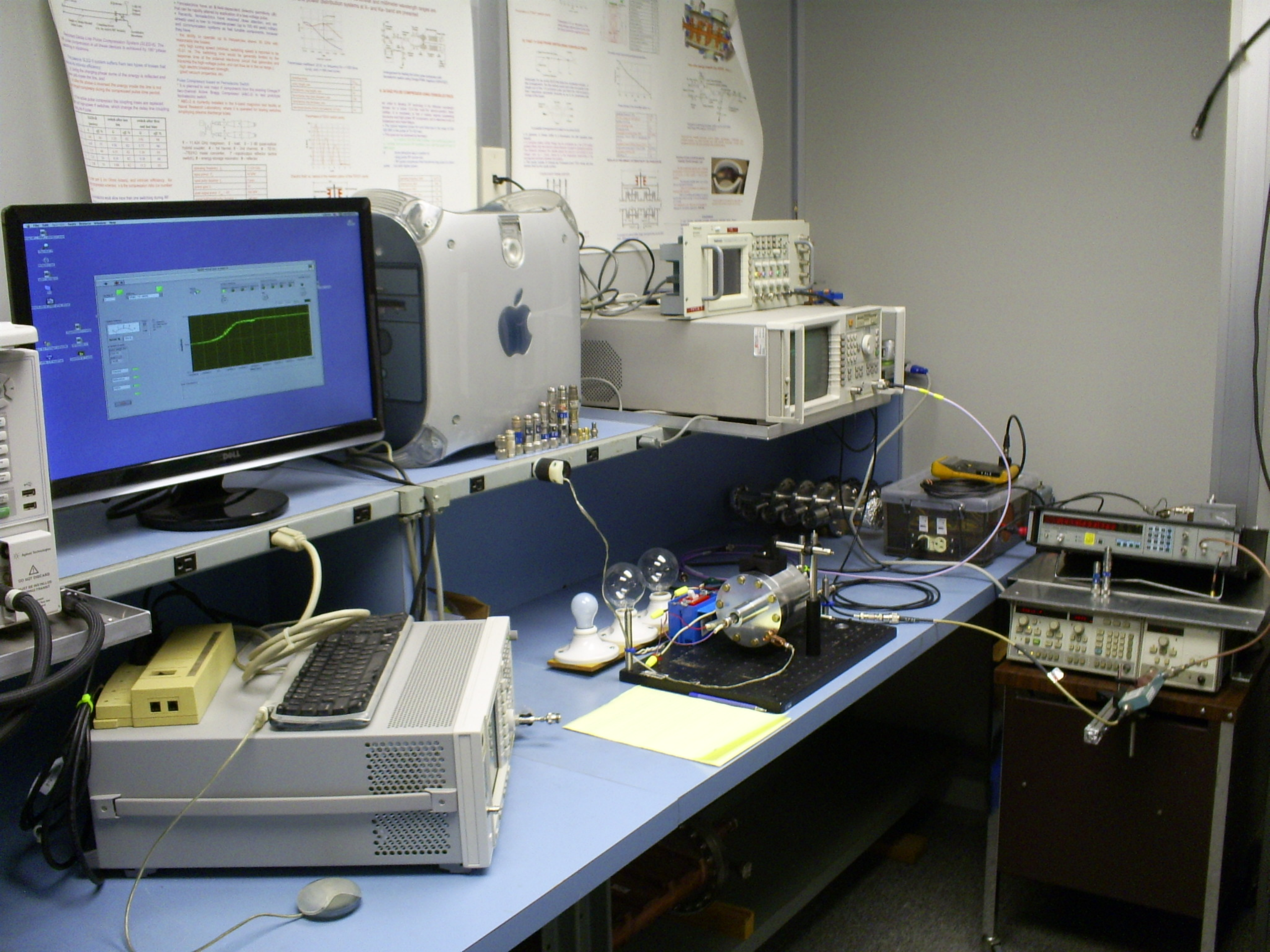}
\end{minipage}
\caption{\textbf{(left)} the schematic of the experimental setup used in the proof-of-principle experiments. \textbf{(Right)} Photo of the setup.}
\label{fig:1}
\end{figure}

\begin{figure}[htb]
\centering
\begin{minipage}{3.25in}
\centering
\includegraphics[width=3.25in]{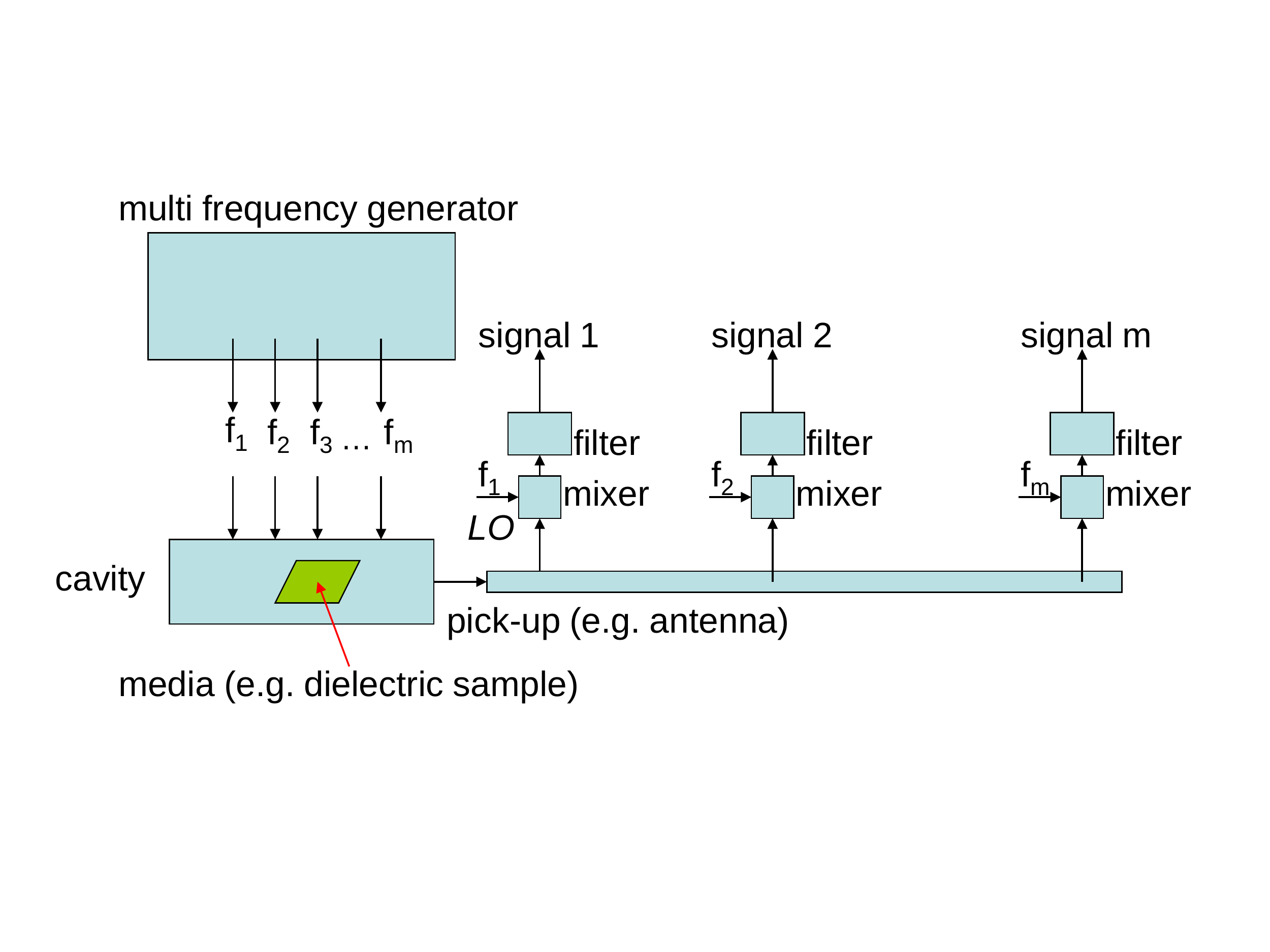}
\end{minipage}
\begin{minipage}{2.5in}
\centering
\includegraphics[width=2.5in]{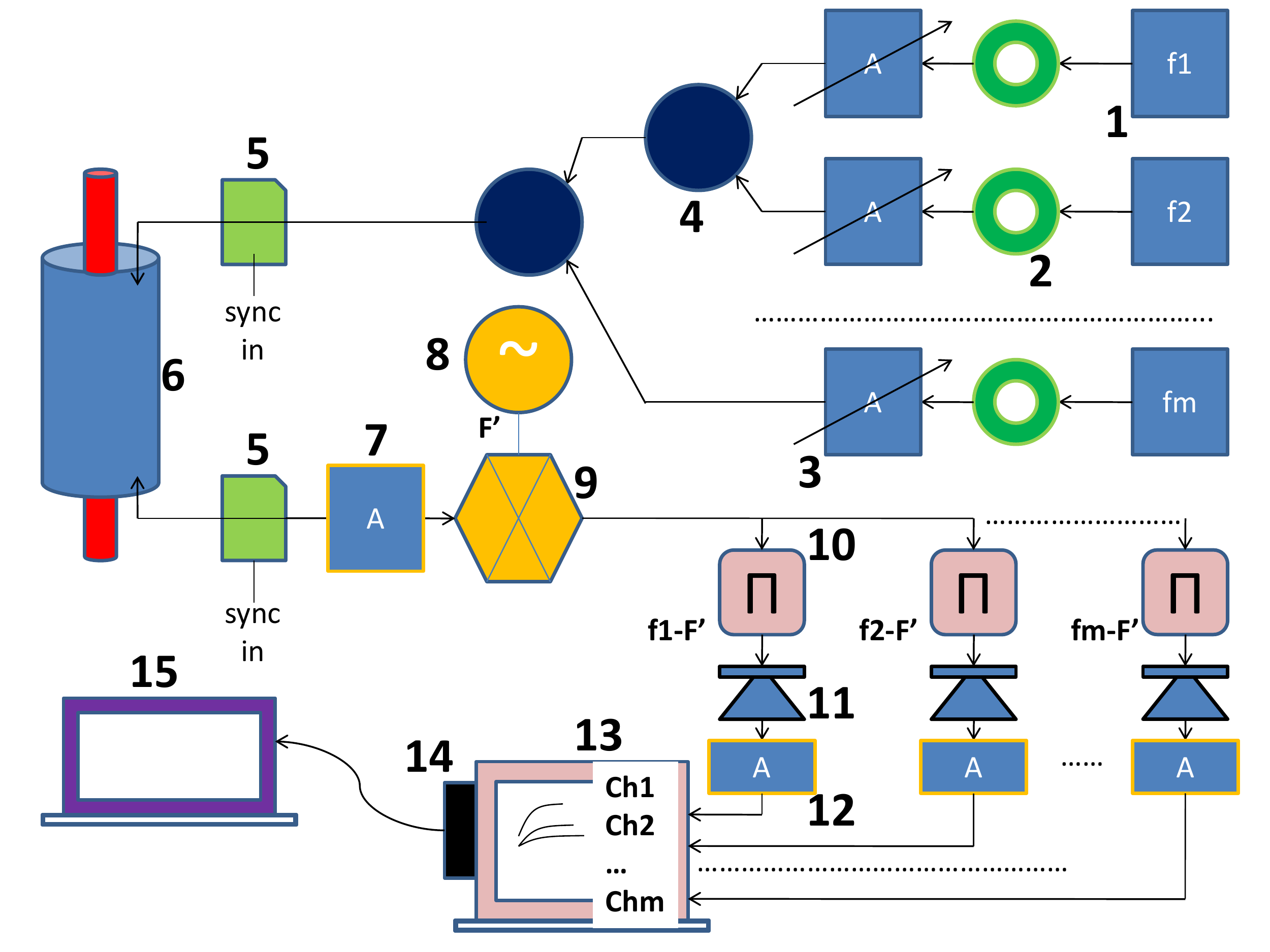}
\end{minipage}
\caption{\textbf{(left)} the principal diagram of generic measurement system, whose particular implementations are as shown in Fig~\ref{fig:1} (left) and in this picture on the right. \textbf{(Right)} Diagram of a detector to install at an accelerator: 1-- tunable frequency generators; 2-- circulators; 3–- variable amplifier; 4–- power combiners; 5–- fast RF switches; 6-– cavity with samples under test; 7-– high frequency output amplifier; 8-– local oscillator; 9-– mixer; 10-– band-pass frequency filters; 11-– crystal detectors; 12-- low frequency amplifiers; 13–- oscilloscope; 14-– GPIB/USB-Ethernet/USB adapter; 15-- computer.}
\label{fig:2}
\end{figure}

The schematic shown in Fig~\ref{fig:1} (left) is a just particular implementation of the principal measurement diagram (Fig~\ref{fig:2}, left) that follows the principles described in (step 2) in the former paragraph. 

Another particular implementation of the generic diagram (Fig~\ref{fig:2}, left) is shown in Fig~\ref{fig:2} (right). This one is envisioned to be used when working with dielectric tubing samples used for accelerator applications and thereby exposed to electron beams and intense wakefields; it includes a fast microwave switch that can suppress radiation emitted by the passing relativistic charge bunch~\cite{ref8}

\begin{figure}[htb]
\centering
\begin{minipage}{3.75in}
\centering
\includegraphics[width=3.75in]{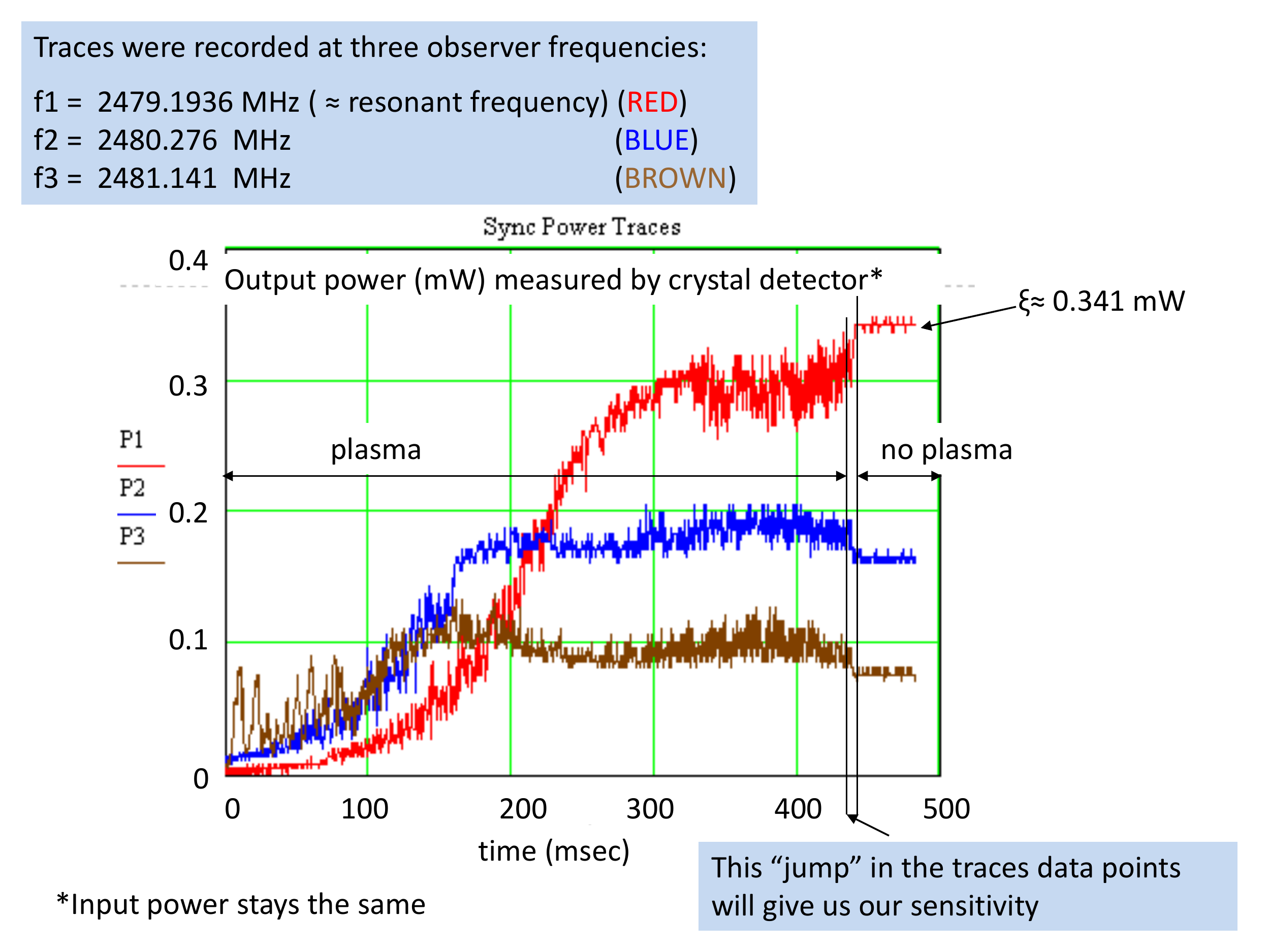}
\end{minipage}
\begin{minipage}{2in}
\centering
\includegraphics[width=2in]{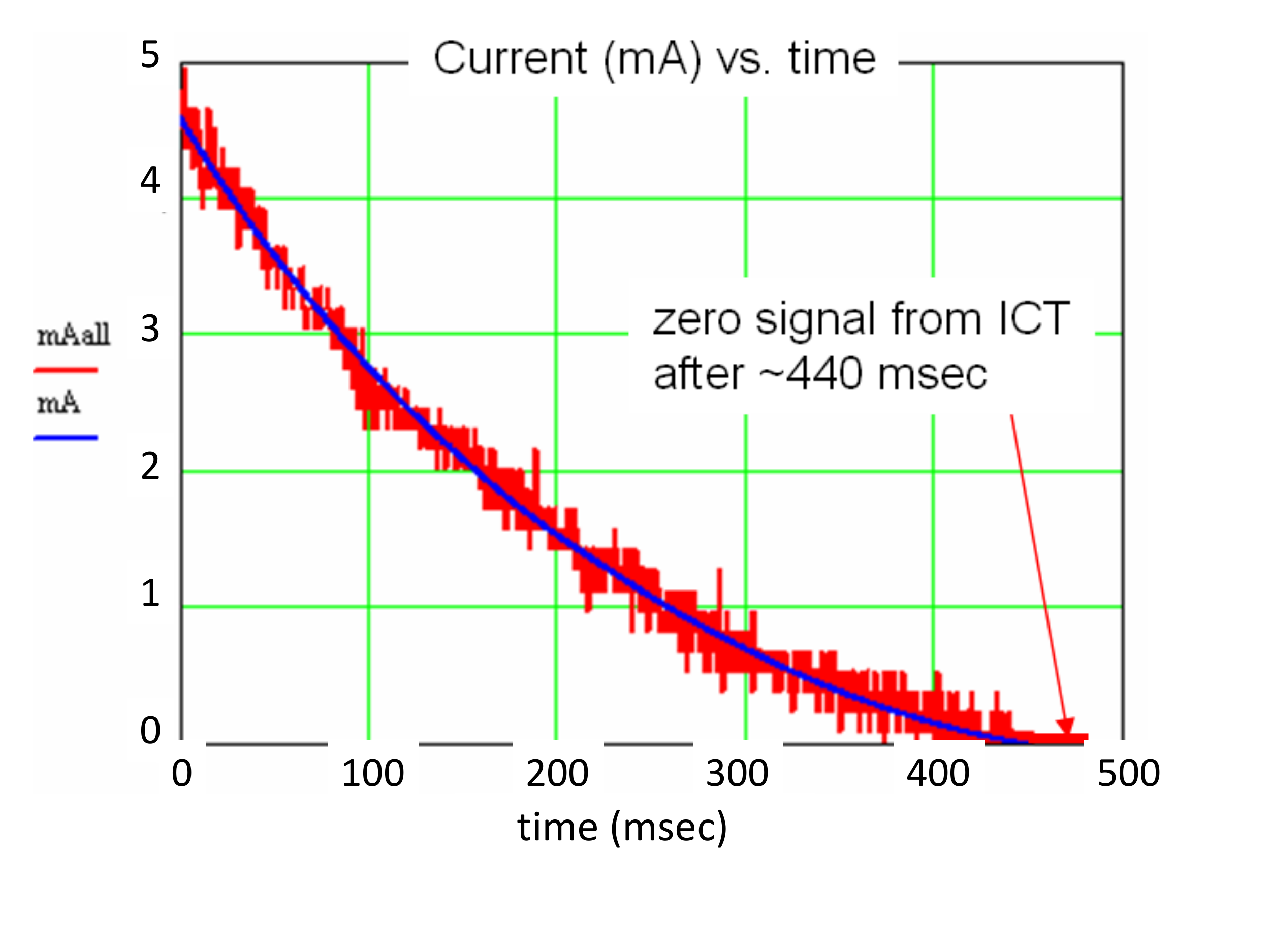}
\end{minipage}
\caption{\textbf{(left)} Microwave signals (0.1mW/div) transmitted through the resonator which is loaded with plasma, using three different frequencies. \textbf{(Right)} Current (1mA/div) behavior during the time (100msec/div) of discharge. }
\label{fig:3}
\end{figure}

An example of recorded traces vs. the time, as obtained in our proof-of-principle runs with plasma-discharge columns is shown in Fig~\ref{fig:3} (left). The corresponding current behavior (which approximately tracks the electron concentration) is shown in Fig~\ref{fig:3} (right). If $\xi$ is the ratio (linear) of the transmitted to incident power in the absence of the discharge (at the resonant frequency F$_{\xi}$) , T$_{m}$ is the ratio (of the transmitted to incident power) at the observer frequency fm in the presence of the discharge (time-dependent value), and F and Q, are the change of resonant frequency and Q-factor respectively, then one finds that:

\begin{equation}
F=\frac{f _{m}+f _{k}}{2}+\frac{F _{\xi}^{2}}{f _{m}-f _{k}}\frac{T _{m}-T _{k}}{\xi}\frac{\xi ^{2}}{T _{m}T _{k}}\frac{1}{8Q _{\xi}^{2}}
\end{equation}

\begin{equation}
\frac{1}{Q^{2}}=\frac{\xi}{T _{k}}\frac{1}{Q _{\xi}^{2}}-\bigg(\frac{f _{m}-f _{k}}{F _{\xi}}+\frac{F _{\xi}}{f _{m}-f _{k}}\frac{T _{m}-T _{k}}{\xi}\frac{\xi ^{2}}{T _{m}T _{k}}\frac{1}{4Q _{\xi}^{2}}\bigg)^{2}
\end{equation}

\begin{figure}[htb]
\centering
\begin{minipage}{2.9in}
\centering
\includegraphics[width=2.9in]{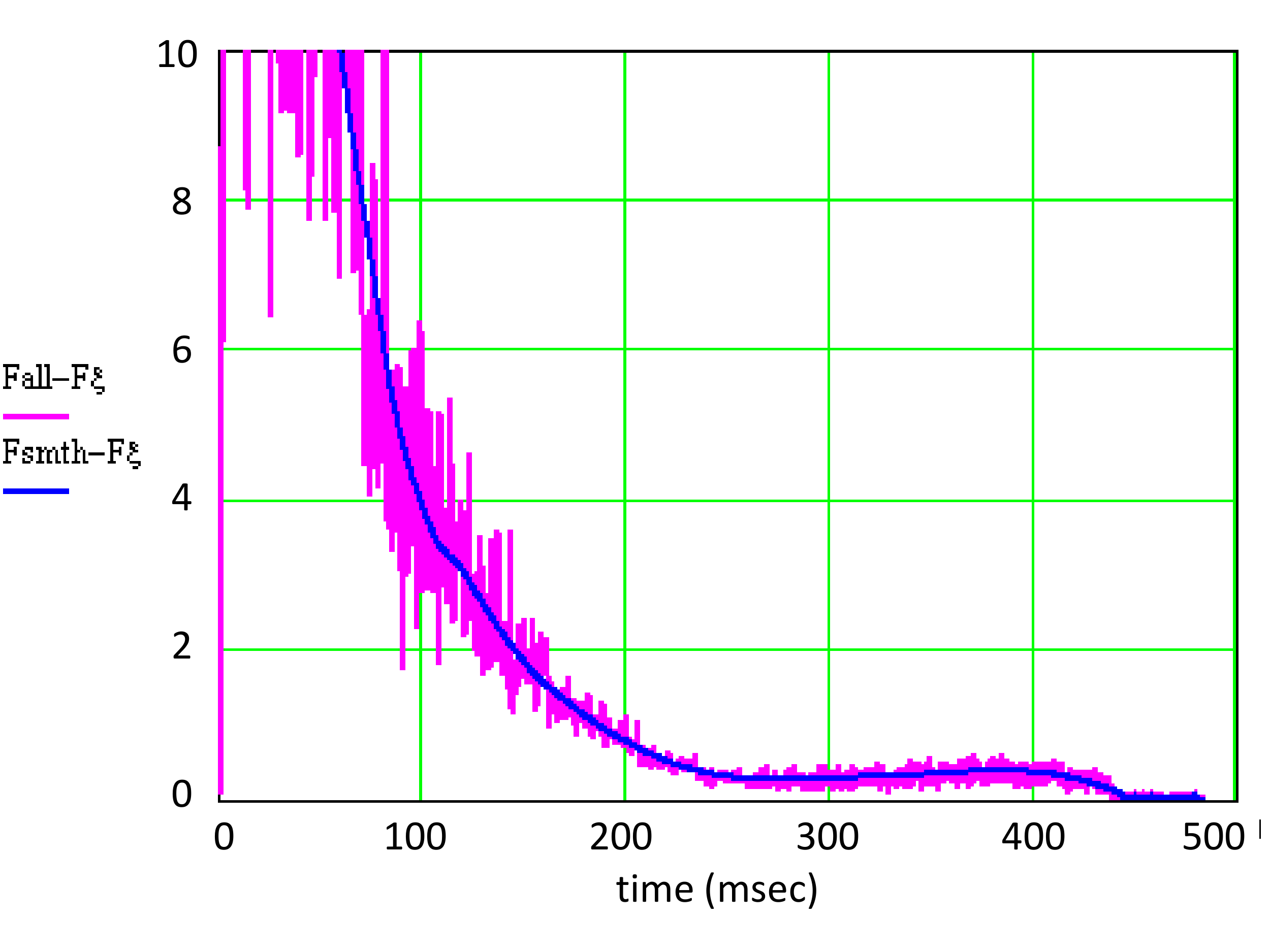}
\end{minipage}
\begin{minipage}{2.9in}
\centering
\includegraphics[width=2.9in]{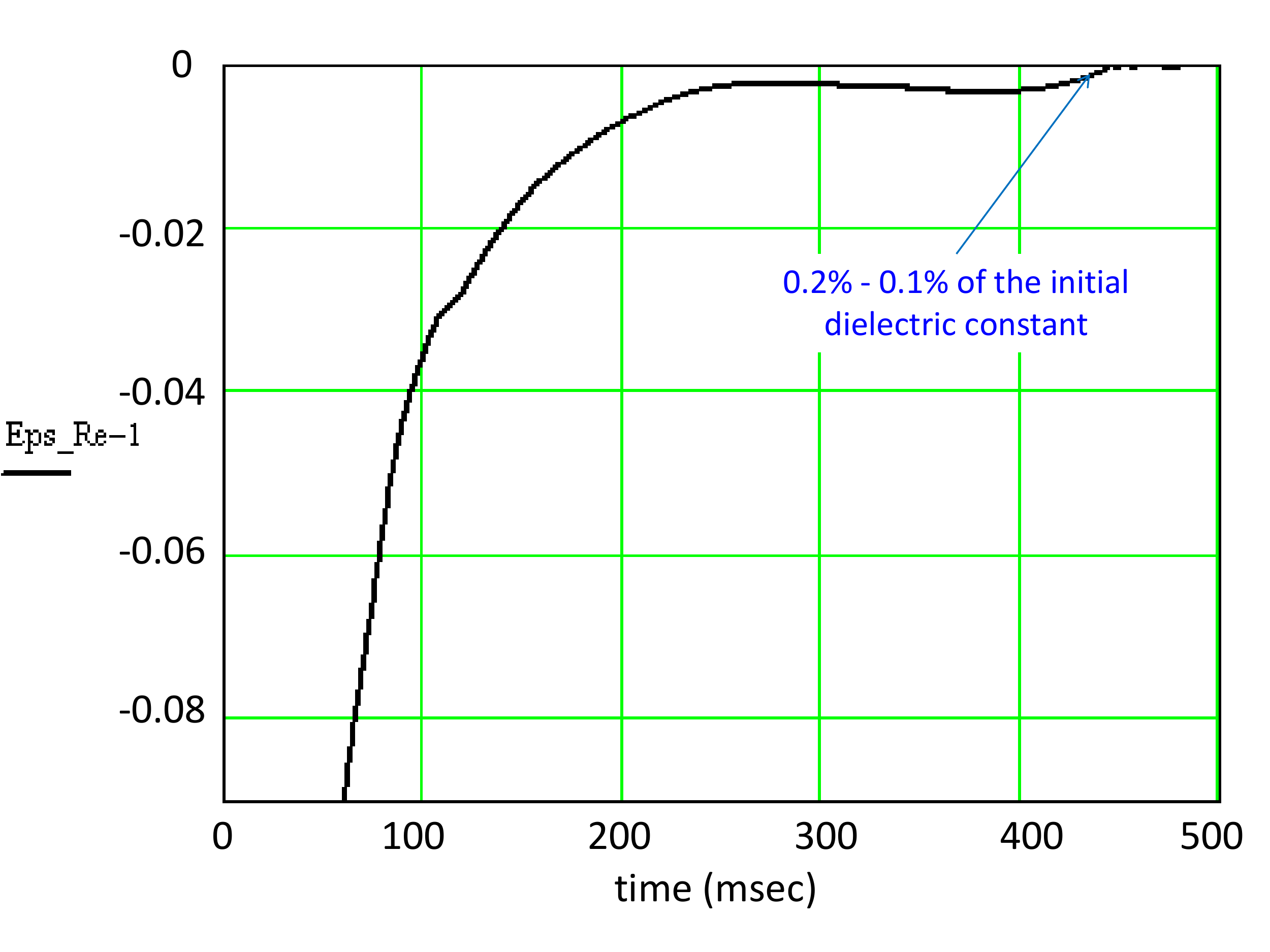}
\end{minipage}
\caption{\textbf{(left)} Changes in the resonant Frequency (2MHz/div) vs. time (100msec/div). \textbf{(Right)} Changes in real part of dielectric constant (0.02/div) vs. time (100msec/div)}
\label{fig:4}
\end{figure}

One sees that at least two external frequencies are required to find the changing resonant frequency and Q-factor. Having a few observer frequencies allows one to have an effective means to filter noise and reject erroneous signal values. Figure~\ref{fig:4} (left) shows an example of the resulting changes in the cavity resonant frequency. Similarly one finds the changes in the Q-factor. 

\begin{figure}[htb]
\centering
\begin{minipage}{2.9in}
\centering
\includegraphics[width=2.9in]{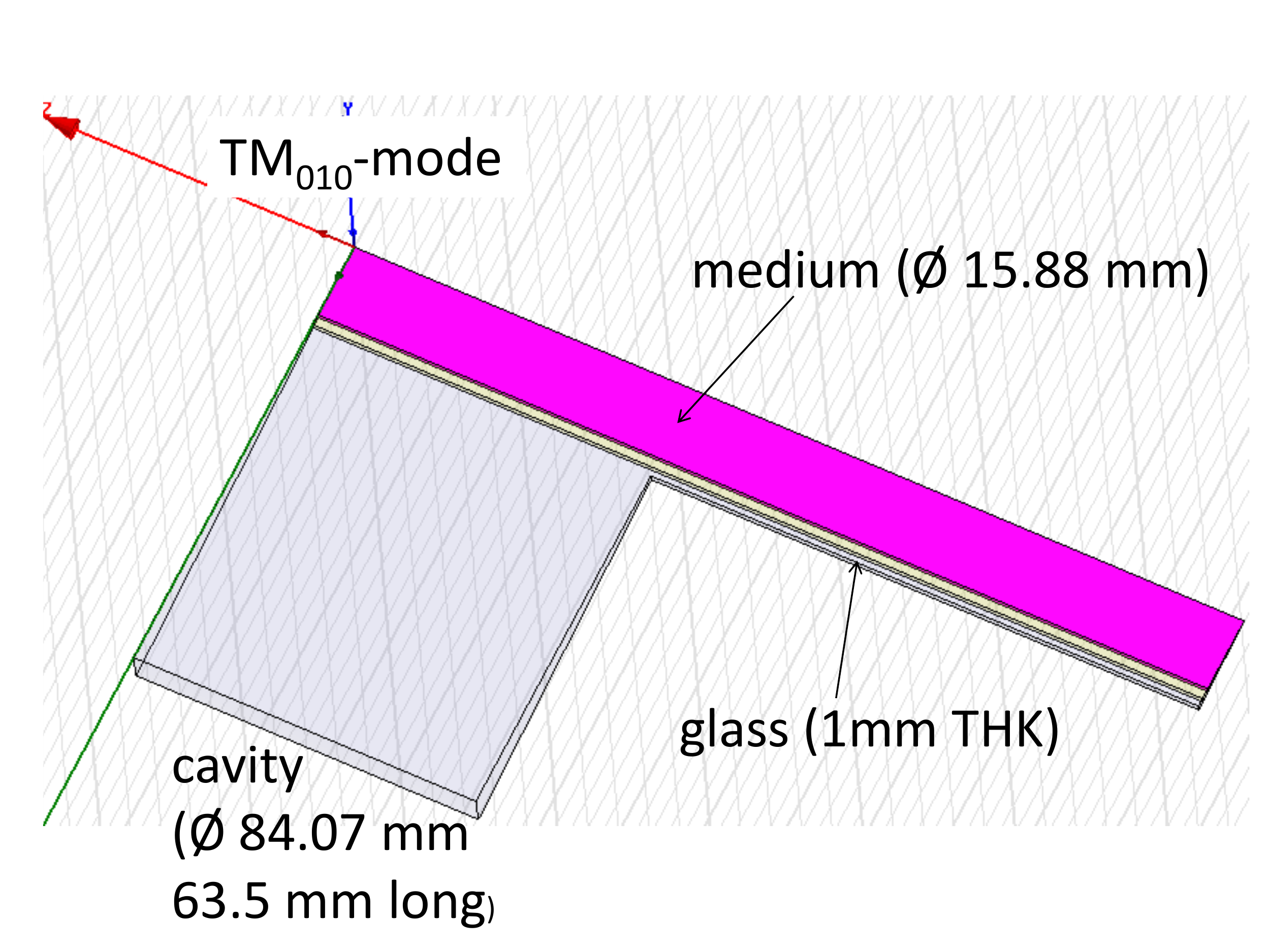}
\end{minipage}
\begin{minipage}{2.9in}
\centering
\includegraphics[width=2.9in]{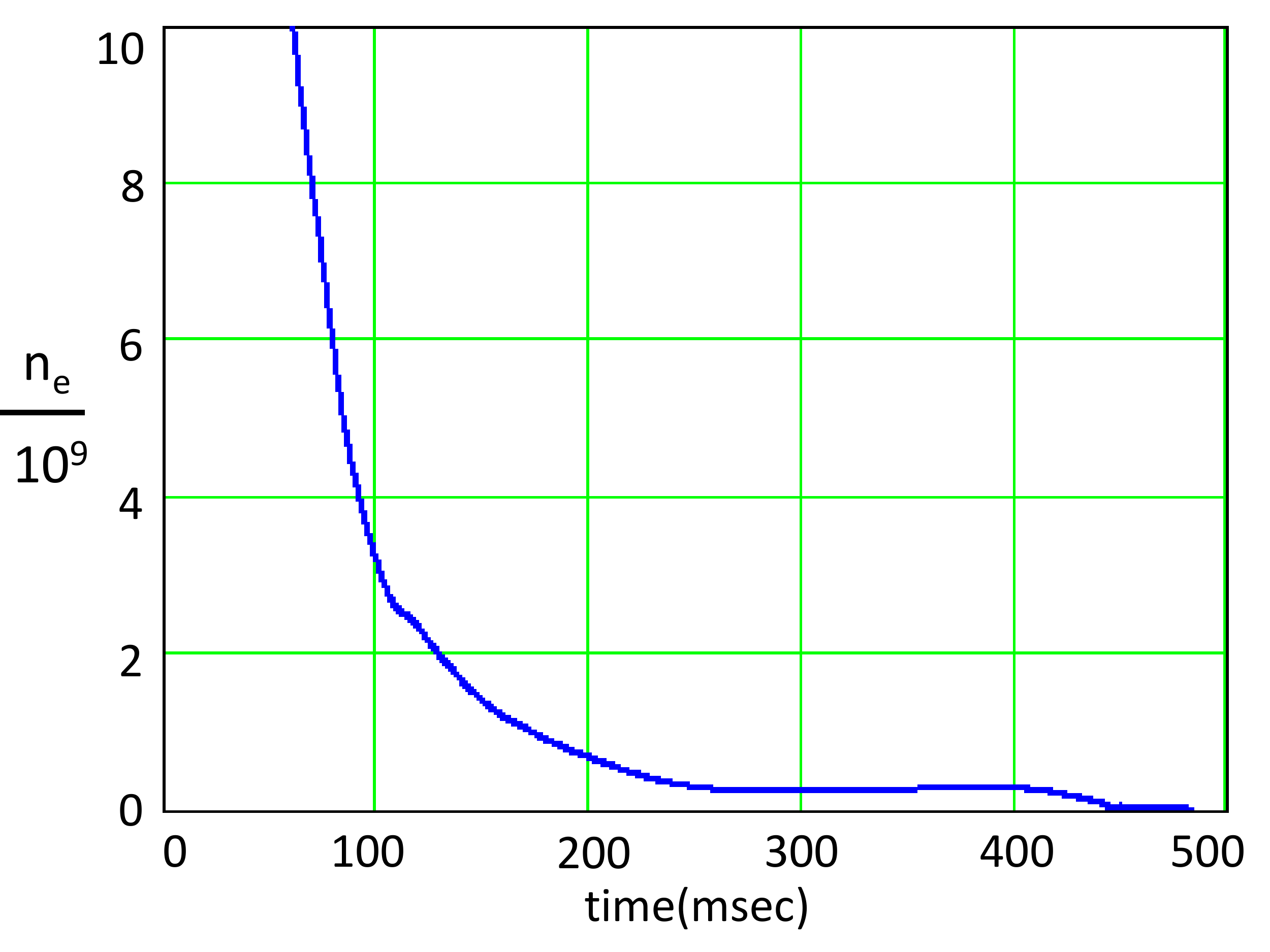}
\end{minipage}
\caption{\textbf{(left)} geometry of the used sample. \textbf{(Right)} Electron density (2$\times$10$^{9}$ cm$^{-3}$/div) vs. time (100msec/div)}
\label{fig:5}
\end{figure}

To find the changes in the sample dielectric constant and loss-tan as per (step 4) one uses:

\begin{equation}
\frac{\Delta\epsilon}{\epsilon}=1+A\frac{F-F _{\xi}}{F _{\xi}}+B\bigg(\frac{F-F _{\xi}}{F _{\xi}}\bigg)^{2}+...
\end{equation}

\begin{equation}
\Delta(losstan)=C\bigg(\frac{1}{Q}-\frac{1}{Q _{\xi}}\bigg)\bigg(1+D\frac{F-F _{\xi}}{F _{\xi}}+E\bigg(\frac{F-F _{\xi}}{F _{\xi}}\bigg)^{2}+...\bigg)
\end{equation}

where the coefficient $A,B, C,… E$ are geometry-dependent and are found after numerical simulations in e.g. HFSS. An example of the found changes in the dielectric constant is shown in Fig.4 (right). The coefficient A, …, E were found to be -22.558, -5.4×10-4, 11.305, 22.581, and 648.776 respectively for the geometry as shown in Figure~\ref{fig:5} (left). 

To find corresponding changes in the density of free electrons, we will use a reasonable assumption about our medium under test. For instance, for the case described here, we can use well known relationships between the density of free electrons, n$_{e}$, and their collision frequency, F$_{coll}$ and the plasma dielectric constant $\epsilon _{0}$($\epsilon _{Re}$ + j$\epsilon _{Im}$).

\begin{equation}
\epsilon _{Re}=1-\frac{n _{e}\cdot e^{2}}{m_{e}\epsilon_{0}(2\pi F_{\xi})^{2}}\frac{1}{1+(F_{coll}/F_{\xi})^{2}}
\end{equation}

\begin{equation}
\epsilon _{Im}=\frac{n _{e}\cdot e^{2}}{m_{e}\epsilon_{0}(2\pi F_{\xi})^{2}}\frac{F_{coll}/F_{\xi}}{1+(F_{coll}/F_{\xi})^{2}}
\end{equation}

where $e$, and  $m_{e}$ are the electron charge and mass respectively, and $\epsilon _{Im}$ = $\epsilon _{Re}\cdot loss-tan$.  Knowing $\epsilon _{Re}$ and $loss-tan$ and solving the equations above one can find the free electron density and collision frequency. In the case of the dielectric thin-wall tubing, the model might be different (it is not a scope of this article to describe it) but the principles remain the same.

Figure 5(right) gives an example of the resulting free electron density. Observe that the detected minimum electron density is about 10$^{8}$ cm$^{-3}$, and the method dynamic range is $\sim$20dB.

\section{Discussion}
The upper limit is dictated here by the limited number of observer frequencies (only 3). If there is a high density of free electrons, the frequency shift is significant, and is out of the coverage range by the comb of the observer frequencies: this brings the signal-to-noise ratio down and beyond detection. If however, the frequency comb can cover a wide range, the dynamic range can be increased (in principle by tens of dB). 
The lower electron density limit can be improved by increasing the cavity Q-factor. This follows from the fact that, at low densities of free electrons, n$_{e}$ $\sim$ BW$_{\xi}$ . In our present setup, Q $\approx$1000, and the minimum density n$_{e, min}\sim$ ~ 10$^{8}$ cm$^{-3}$; If one works, however, with Q $\sim$ 3000-5000, this will still allow us to capture the processes on the micro-sec time-scale, yet to detect n$_{e, min} \sim$ 2$\div$4 $\times$ 10$^{7}$ cm$^{-3}$.

\end{document}